\documentclass[useAMS,usenatbib]{mnras}
\usepackage[latin1]{inputenc}
\usepackage{amsmath}
\usepackage{amsfonts}
\usepackage{amssymb}
\usepackage{graphicx}
\usepackage{epstopdf}

\newcommand{\dd}{{\rm d}}
\newcommand{\xx}{\mathbfit{x}}
\newcommand{\vv}{\mathbfit{v}}

\newcommand{\tcross}{\ensuremath{t_{\rm char}}}
\newcommand{\trelax}{\ensuremath{t_{\rm relax}}}
\newcommand{\e}{\ensuremath{\bar\epsilon}}

\title{On the probabilistic approach to the N-body problem}

\author[M.~Romero and Y.~Ascasibar]
{
M.~Romero$^{1}$ and Y.~Ascasibar$^{1,2}$ \\
$^{1}$ Departamento de F\'{i}sica Te\'{o}rica, Universidad Aut\'{o}noma de Madrid, Madrid 28049, Spain\\
$^{2}$ Astro-UAM, UAM, Unidad Asociada CSIC
}

\pagerange{\pageref{firstpage}--\pageref{lastpage}}
\pubyear{2017}

\begin{document}

\maketitle
\label{firstpage}

\begin{abstract}
{This work discusses the main analogies and differences between the deterministic approach underlying most cosmological N-body simulations and the probabilistic interpretation of the problem that is often considered in mathematics and statistical mechanics.
In practice, we advocate for averaging over an ensemble of $S$ independent simulations with $N$ particles each
in order to study the evolution of the one-point probability density $\Psi$ of finding a particle at a given location of phase space $(\xx,\vv)$ at time $t$}.
The proposed approach is extremely efficient from a computational point of view, with modest CPU and memory requirements, and it provides {an} alternative to traditional N-body simulations when the goal is to study the average properties of N-body systems, at the cost of abandoning the notion of well-defined trajectories for each individual particle.
{In one spatial dimension}, our results, fully consistent with those previously reported in the literature for the standard deterministic formulation of the problem, highlight the differences between the evolution of the one-point probability density $\Psi(x,v,t)$ and the predictions of the collisionless Boltzmann (Vlasov-Poisson) equation, as well as the relatively subtle dependence on the actual finite number $N$ of particles in the system.
{We argue that understanding this dependence with $N$ may actually shed more light on the dynamics of real astrophysical systems than the limit $N\to\infty$}.
\end{abstract}


\section{Introduction}
\label{sec_introduction}

At large scales, almost all interactions in the Universe are dominated by gravity.
Most of the processes studied in astrophysics and cosmology involve, one way or another, a set of objects (planets, stars, galaxies...) that interact with each other through their mutual gravitational attraction.
More specifically, the N-body problem refers to the evolution of a self-gravitating system formed by $N$ point-like particles, and its solution is of vital importance in many contexts, from gravitational instability (e.g. planet, star or galaxy formation) to the dynamics of systems in equilibrium (which determines the internal properties of planetary systems, star clusters, galaxies, galaxy clusters...).

Writing the equations of motion for an N-body system is easy, but solving them is not.
Analytical solutions are only known for $N=2$ and some particular conditions with $N=3$.
Thus, in most applications of interest in astrophysics, the equations are solved numerically, using a set of initial conditions and studying the evolution of the system in time.
When $N$ is small, the mutual interactions between particles can be evaluated directly, and the main source of uncertainty is the numerical integration error.
Nevertheless, the number of force evaluations grows as $N^2$, and the direct method quickly becomes impractical.
Using hybrid parallelisation techniques in specialized (CPU+GPU) hardware, the maximum number of particles that can be dealt by direct integration is of the order of one million \citep{Wang+15}.

For larger values of $N$, it is necessary to use approximations that reduce the number of force evaluations.
For example, by computing the accelerations from close neighbours directly, whereas the interaction between groups of particles far away are obtained from their mass moments.
Many of these algorithms use a tree structure to choose what particles are close and which are not, hence the number of evaluations scales as $N \log(N)$ \citep{Barnes&Hut86} or even close to $N$ \citep{Dehnen00}.
For the reader interested in this particular topic, see e.g \cite{Aarseth03}.

In many cases, though, the number of objects to simulate in a real system is much greater than the number of particles that the best supercomputers can handle.
The Milky Way, for example, contains approximately $10^{11}$ stars and about $10^{12}$ $M_{\odot}/m_{dm}$ particles of dark matter with mass $m_{dm}$ \citep{Binney&Tremaine08},
and thus it is virtually impossible to simulate every one of them individually.
When the number of particles is so large, one usually relies on the long-range character of the gravitational interaction to make an important approximation: the gravitational acceleration of any particle is not dominated by its nearest neighbours, but by the mean potential created by the global density distribution.
Under these conditions (i.e. in the limit $N \rightarrow \infty$), it is assumed that the system is fully described by a continuous \emph{distribution function}
\begin{equation}
f(\xx,\vv) \equiv \frac{ \dd M }{ \dd^3\xx\,\dd^3\vv }
\label{eq_DistrFunctionDef}
\end{equation}
that indicates the density in phase space (i.e. the amount of mass located at position $\xx$ moving with velocity $\vv$ per unit of six-dimensional volume $\dd^3\xx\,\dd^3\vv$).

Given the distribution function, the mass density at any spatial location $\xx$ can be obtained by integrating over all possible velocities
\begin{equation}
\rho(\xx) \equiv \int_{-\infty}^{\infty} f(\xx,\vv)\ \dd^3 v,
\end{equation}
whereas integration over configuration space (the three-dimensional spatial volume) yields the velocity distribution of the system
\begin{equation}
\eta(\vv) \equiv \int_{-\infty}^{\infty} f(\xx,\vv)\ \dd^3 x.
\end{equation}
The temporal evolution of the distribution function is given by the collisionless Boltzmann equation
\begin{equation}
\frac{\dd f}{\dd t} = \frac{\partial f}{\partial t} + \vv \frac{\partial f}{\partial \xx} + \mathbf{a} \frac{\partial f}{\partial \vv} = 0,
\label{eq_LiouvilleTheorem}
\end{equation}
which is nothing other than Liouville's theorem (density conservation in phase space), where $\mathbf{a}(\xx)$ corresponds to the acceleration associated to the Newtonian gravitational potential\footnote{In this case, expression~\eqref{eq_LiouvilleTheorem} is also called Vlasov-Poisson equation \citep[see][for a discussion of the optimal nomenclature]{Henon82}.}.

Traditionally, it is considered that equation~\eqref{eq_LiouvilleTheorem} provides a valid approximation to the dynamics of a \emph{finite} N-body system as long as the individual interactions between two or more particles may be neglected.
The range of validity is usually expressed in terms of the relaxation time
\begin{equation}
\trelax \sim 0.1 \frac{N}{\ln(N)} \tcross,
\label{eq_RelaxTime3D}
\end{equation}
where $\tcross$ denotes the time it takes for a particle to cross (or take half an orbit around) the system.
The relaxation time $\trelax$ is also meant to provide an idea of the time it takes a particle to randomize its velocity due to individual encounters (collisions) with other bodies and lose all memory of its initial conditions.
The exact value of $\trelax$ varies greatly depending on the local density and the total number of particles.
For the stars in the Milky Way, $N \sim 10^{11}$ and $\tcross \sim 0.1$~Gyr; $\trelax$ is much longer than the age of the Universe, and therefore~\eqref{eq_LiouvilleTheorem} is considered a good approximation.
On the other hand, for globular ($\tcross \sim 0.1$~Myr, $N \sim 10^5$) and open ($\tcross \sim 1$~Myr, $N \sim 100$) clusters, $\trelax \sim 87$ and 2~Myr, respectively, and two-body encounters play an important role in the evolution of the distribution function \citep{Binney&Tremaine08}.

When $N$ is small (and/or $\trelax$ is smaller than the time scales of interest), interactions are simulated directly.
When $N \rightarrow \infty$ and $\trelax \rightarrow \infty$, equation~\eqref{eq_LiouvilleTheorem} can be integrated in many ways \citep[see e.g.][and references therein, for a comparison of different methods]{Colombi&Touma14, Colombi+15, Mocz&Succi17}, but a lot of them rely on sampling the distribution function, assumed to be approximately continuous, with a finite number of tracers, ${N_{\rm tracers}} \ll N$, that represent a certain amount of mass.
The continuous function $f(\xx,\vv,t)$ is approximated by means of these tracers, interacting with each other through their mutual gravity, with some kind of small-scale filtering to prevent two-body interactions \citep[because they no longer represent point-like masses, as in the direct case; see e.g.][for an in-depth discussion of this `gravitational softening']{Dehnen01}.
It is (implicitly or explicitly) assumed that, the higher the number of tracers used, the more the solution will resemble~\eqref{eq_LiouvilleTheorem}, reaching it asymptotically as ${N_{\rm tracers}}\to\infty$.
{In practice, a set of statistical tests show that the results of high-resolution N-body simulations of a \citet{Henon64} sphere\footnote{{Constant density up to a certain radius, Maxwellian velocity distribution.}} are consistent with local Poisson sampling of the distribution function estimated by a Vlasov code \citep{Colombi+15}.
However, the agreement between both methods worsens as the velocity dispersion becomes smaller, and convergence in the limit ${N_{\rm tracers}}\to\infty$ is as yet unproven.}

The main aim of the present work is to {advocate for a completely different route.
Rather than focusing on the limit $N\to\infty$, we argue that, in order to understand the dynamics of real systems}, it is often not important to know exactly the positions and velocities of all particles\footnote{Something that can hardly be achieved in a traditional N-body simulation anyway, due to the approximations involved in the force evaluation.}, but only the average statistical properties of the distribution function.
The fundamental quantity is the
one-point \emph{probability density} $\Psi(\xx,\vv,t)$ of finding a particle with coordinates $(\xx,\vv)$ at time $t$.
We are only interested in the evolution of this function, and \emph{not} on the trajectories of individual particles, whose coordinates at any time -- including $t=0$ -- are assumed to be $N$ statistically-independent \emph{random} realisations of $\Psi(\xx,\vv,t)$, at variance with the traditional, deterministic formulation of the problem.

The evolution of such an intrinsically stochastic N-body system, fully specified by the initial condition $\Psi(\xx,\vv,0)$, will be estimated by averaging over several N-body simulations, and the results will be compared with the collisionless Boltzmann equation~\eqref{eq_LiouvilleTheorem}.
{The concept of a statistical ensemble is certainly not new, and it has previously been used to derive many analytical results \citep[see e.g.][]{Binney&Tremaine08, Campa+09, Chavanis13, Colombi+15}, but numerical applications \citep[e.g.][]{JW10} are much more scarce, especially within the context of astrophysics and cosmology.
Here we would like to advocate for this procedure on both theoretical and practical grounds.
On the other hand, we would also like to stress that, even if they are closely related, the probabilistic and the deterministic interpretation of the N-body problem differ in several crucial respects, and a careful comparison between them is necessary in order to obtain robust conclusions about real astrophysical systems.
}

We will first discuss the distribution function and the collisionless Boltzmann equation in more detail, including some analytical solutions, in Section~\ref{sec_SomeSolutions}.
The main aspects of {the} probabilistic approach are then presented in Section~\ref{sec_ProbabilityDensity}, highlighting the similarities and differences between $f$ and {the \emph{expected} phase-space density $\bar f(\xx,\vv,t) = M\Psi(\xx,\vv,t)$, where $M$ is the total mass of the system}.
The evolution of the probability density $\Psi$ is investigated by a series of numerical experiments, described in Section~\ref{sec_NBodySimulations}.
The main results are shown in Section~\ref{sec_Results}, and they are compared with {their deterministic analogues} in Section~\ref{sec_Discussion}.
Our conclusions are briefly summarized in Section~\ref{sec_Conclusions}.

\section{Analytical solutions for the distribution function}
\label{sec_SomeSolutions}

Let us start by considering the evolution of a truly continuous fluid in phase-space, stressing that this is \emph{different} from an N-body system, even as $N\to\infty$ (see below).
For the sake of simplicity, as well as computational accuracy and efficiency, we will work in one spatial dimension {with plane-parallel symmetry}.
The Poisson equation
\begin{equation}
\frac{\partial^2 \phi(x)}{\partial x^2} = {4}\pi G \rho (x)
\label{eq_PoissonEq}
\end{equation}
implies that the gravitational potential at any spatial location $x$ is given by the expression
\begin{equation}
\phi(x) = 2\pi G \int_{-\infty}^{\infty} \rho(x')\ |x-x'|~dx',
\label{eq_potential_rho}
\end{equation}
and the gravitational acceleration that appears in the collisionless Boltzmann equation
\begin{equation}
a(x) 
= 2\pi G \left[ \int_{x}^{\infty} \rho(x')~dx' - \int_{-\infty}^{x} \rho(x')~dx' \right]
\label{eq_acceleration_rho}
\end{equation}
is just proportional to the difference between {total mass surface density} to the right and the left of $x$.

Equations~\eqref{eq_LiouvilleTheorem} and~\eqref{eq_acceleration_rho} provide an \emph{exact} description of the evolution of the distribution function $f(x,v,t)$ for a continuous one-dimensional fluid.
In general, it is necessary to solve these equations by numerical integration, but analytical solutions may be found for some particular sets of initial conditions.
Here we will make use of two such solutions to illustrate the similarities and differences between the distribution function $f$ and the probability density $\Psi$.

{Finding an analytic solution for the evolution of a system of $N$ self-gravitating bodies} is even more difficult than for the continuous fluid, but, once again, it is possible for very specific initial conditions.
For a discrete distribution of $N$ {infinite plane sheets with equal mass surface density $\sigma$ (hereafter referred to as `particles')}, the phase-space density is a sum of Dirac delta functions
\begin{equation}
 f(x,v,t) = \sigma \sum_{i=1}^N \delta(x-x_i(t))\ \delta(v-v_i(t)),
 \label{eq_sum_deltas}
\end{equation}
and the gravitational acceleration can be expressed as
\begin{equation}
a(x) = 2\pi G \sigma \left[ N_{+}(x) - N_{-}(x) \right]
\label{eq_TrueAcc}
\end{equation}
where $N_{+}$ and $N_{-}$ denote the number of particles to the right and left of $x$, respectively.

{Although it is possible define some kind of mean field limit as $N\to\infty$ \citep[see e.g.][]{Campa+09, Chavanis13}, equation~\eqref{eq_sum_deltas} will never become a continuous distribution, no matter how large $N$ may be.
Even a (countable) infinite sum of deltas is fundamentally different from a continuous function $\mathbb{R}^2\to\mathbb{R}$ that provides a \emph{real} density for any possible location in the two-dimensional phase space of positions and velocities.
Even after coarse-graining, any unweighted integral of equation~\eqref{eq_sum_deltas} over a finite phase-space volume $V$ can only provide an integer multiple of $\sigma\to 0$, and there will always be some space, no matter how small\footnote{{In fact, $|\mathbb{R}|<|\mathbb{Q}|$ implies that, in a mathematical sense, \emph{almost all} real numbers are irrational.}}, between one particle and the next.

Nevertheless}, all of the considerations above regarding continuous fluids still hold.
In particular, the evolution of $f(x,v,t)$ {for a system composed of a finite number $N$ of discrete particles} is \emph{exactly} determined by the Vlasov-Poisson equation\footnote{{When the distribution function is a sum of Dirac deltas, the conservation equation is usually referred to as the Klimontovich equation.}} at all times.
It is only when one introduces some (often loosely defined) {integration over a certain volume of phase space} in order to obtain a finite and continuous `coarse-grained' distribution function {$\hat f$} that equation~\eqref{eq_LiouvilleTheorem} becomes an approximation, {and particle correlations start to play a role on the evolution of $\hat f$, yielding the Lenard-Balescu equation and the infinite BBGKY hierarchy}.

\subsection{Stationary solutions}
\label{ssec_SteadySolution}

It is easy to show \citep[see e.g.][]{Binney&Tremaine08} that any distribution function of the form $f(\epsilon)$, where $\epsilon=\frac{v^2}{2} + \phi(x)$ denotes the energy per unit of mass of a particle located at $(x,v)$, is a stationary solution of the collisionless Boltzmann equation:
\begin{eqnarray}
\frac{\partial f}{\partial t} + v \frac{\partial f}{\partial x} + a \frac{\partial f}{\partial v} &=& \frac{\partial f}{\partial t} + v \frac{\partial \phi}{\partial x}\frac{\partial f}{\partial \epsilon} - v \frac{\partial \phi}{\partial x}\frac{\partial f}{\partial \epsilon} = \nonumber \\
&=& \frac{\partial f}{\partial t}\ =\ 0.
\label{eq_fe_isSteady}
\end{eqnarray}
{A continuous fluid with such a distribution function will remain strictly stationary at all times.
Of course, these solutions are not possible for any system composed of a natural number $N$ of particles}, for the simple reason that~\eqref{eq_sum_deltas} can only adopt the form $f(\epsilon)$ in the trivial case $x_i=v_i=0$ for every particle $i$.
For any other configuration, a discrete system can \emph{never} be stationary in the sense of $\frac{\partial f}{\partial t} = 0${, not even in the limit $N\to\infty$}.

{Nevertheless}, one may expect that, on {very long} time scales, collisions would eventually drive the system towards {a state where the `average' $\hat f$ (over a certain period of time and/or phase-space volume) remains approximately constant}.
Other (collisionless) processes, collectively known as `violent relaxation' \citep{Lynden-Bell67} may achieve a similar effect on much shorter time scales, and several authors \citep[e.g.][among others]{Ogorodnikov57, Lynden-Bell67, Shu78, Plastino&Plastino93, Hjorth&Williams10} have tried to derive the precise function $f(\epsilon)$ from statistical mechanics arguments (e.g. by maximising a suitably-defined entropy).
These quasi-stationary states are out of equilibrium, and it is expected that, after a sufficiently long time (that increases with particle number), collisions will drive them towards thermal equilibrium \citep[see e.g.][and references therein]{Rybicki71, JW10, Levin+14}.

{
In the one-dimensional case, staring from initial conditions very similar to the ones we consider in the present work, it has been found \citep{JW10} that the system first relaxes to a quasi-stationary state in a relatively short time (of the order of tens of dynamical times, roughly independent on the number of particles), and then it stays there for a much longer time, proportional to $N$, until it eventually approaches thermal equilibrium.
Although a precise definition of this time scale is not trivial, it is certainly several orders of magnitude larger than the time required to reach the quasi-stationary regime, and values of the order of $10^6$ dynamical times are reported by \citet{JW10} for $N\sim 100$.
}

\subsection{Periodic solution}
\label{ssec_PeriodicSolution}

A simple analytic solution can be obtained for a continuous system starting at rest ($v=0$) at $t=0$, with uniform spatial density {$\rho(x)=\frac{M}{L^3}$} between $x = -\frac{L}{2}$ and $x=\frac{L}{2}$:
\begin{equation}
 f(x,v,0) = {\frac{M}{L^3}}\ H(x+\frac{L}{2})\ H(\frac{L}{2}-x)\ \delta(v),
 \label{eq_barIC_f}
\end{equation}
where $H(x)$ denotes the Heaviside step function.

Without loss of generality, we will choose from now on our units of mass, length, and time, so that {$M=1$, $L=1$, and $(2\pi G\frac{M}{L^3})^{-1/2} = 1$}, respectively.
In these units, the acceleration at the initial time is simply $a(x) = -2x$, and the system will start shrinking homologously; the velocity is proportional to $x$, and thus the density evolves with time, but maintaining the same (constant) form.
Since there is no crossing between fluid elements, the accelerations remain constant during the collapse, and the trajectories describe a parabola in phase space
\begin{equation}
(x,v,t) = [x_0 (1-t^2), -2 x_0 t, t]
\end{equation}
as a function of the initial position $x_0$.

This description is valid until all trajectories simultaneously reach the point $x=0$ at $t=1$, crossing it with velocity $v = -2x_0$.
When the crossing occurs, the acceleration abruptly changes sign from $a(x,t<1)= -2x_0$ to $a(x,t>1) = 2x_0$.
After that, the system expands until all velocities are zero again at $t=2$, obtaining a distribution $(x,v)=(-x_0,0)$ that is perfectly symmetrical to the original.

Summarizing, we have found a periodic solution with a characteristic crossing time $\tcross=2$.
At any time, the system forms an infinitely-thin, straight line in phase space, and it switches between phases of expansion and contraction.
At $t=(n+1/2)\,\tcross$, all points are located at $x=0$, moving at the highest speed, whereas at $t= n\,\tcross$ the distribution reaches null velocity and maximum spatial extent.
Assuming that we plot $x$ and $v$ as the abscissa and ordinate of a phase-space diagram, respectively, the system starts as a horizontal \emph{bar} that rotates clockwise with a period $T=4$.

This {is the \emph{exact} analytical solution for a continuous fluid, and it also describes the motion of} $N$ \emph{equispaced} particles ($x_{i+1}=x_i+\frac{1}{N-1}$) between $x_1=-\frac{1}{2}$ and $x_N=-\frac{1}{2}$ {with surface density $\sigma=\frac{M}{NL^2}$}.
Note that {this is a very special discrete configuration}.
If the particles were randomly distributed (an arguably more physically-motivated, or at least infinitely more likely situation), the initial acceleration would no longer be \emph{exactly} proportional to $x_i$,
and thus the particles will not reach the origin at the same time.
Any perturbation with respect to a perfectly equispaced distribution will grow in time, and the system will evolve differently depending on the precise details of the initial conditions.
The particles cross each other a different times, and one may conjecture that their interactions should drive the system towards a quasi-stationary state.

\section{Probability formulation of the N-body problem}
\label{sec_ProbabilityDensity}

The main idea that we would like to put forward is that, quite often, one is not interested on the evolution of $N$ particles with (infinitely) precisely known initial positions and velocities $x_i$ and $v_i$.
Instead, we would like to address the problem, {thoroughly discussed in the context of statistical mechanics \citep[see e.g.][for recent reviews]{Campa+09, Levin+14},} of $N$ particles whose initial conditions are randomly drawn from the probability density $\Psi(x,v,0)$.
We {thus} abandon the notion of individual particle trajectories and focus on the \emph{expected} phase-space density
\begin{equation}
\bar{f}(x,v,t) = {\frac{M}{L^2}}\ \Psi(x,v,t)
\label{eq_fbar}
\end{equation}
after time $t$, averaged over all possible random realisations of the initial conditions, where $\Psi(x,v,t)$ is the one-point probability density (i.e. neglecting correlations) for the distribution of particles in phase space.
Although the distribution function $f$ for any \emph{individual} realisation of the initial conditions will be, of course, a sum of Dirac deltas, and the number of particles within a certain interval of $(x,v)$ at time $t$ will always be an integer number, the average density $\bar f$ and the probability density $\Psi$ will be, in general, continuous functions with real values, such that
\begin{equation}
\int \Psi(x,v)\ \dd x\, \dd v =1
\label{eq_PsiNormalization}
\end{equation}
at any time $t$.
Moreover, conservation of probability can be written as
\begin{equation}
\frac{d \Psi}{dt} = \frac{\partial \Psi}{\partial t} + v \frac{\partial \Psi}{\partial x} + a \frac{\partial \Psi}{\partial v} = 0,
\label{eq_PsiConservation}
\end{equation}
which is analogous to the collisionless Boltzmann equation, with the exception that the acceleration $a$ is now a stochastic variable, subject to random statistical fluctuations of $N_+$ and $N_-$ in~\eqref{eq_TrueAcc}.
{Averaging over the statistical ensemble of all possible random realisations of the initial condition $\Psi(x,v,0)$ in order to obtain a fully deterministic, ordinary differential equation for the evolution of $\Psi$ and/or $\bar f$, additional terms would appear on the right hand side due to collisions and correlations between particles (the BBGKY hierarchy), and different approximations may be used in order to close the system \citep[see e.g.][and references therein]{Campa+09, Chavanis13}}.

Here we adopt a more practical approach, simply aiming for an approximate solution {of~\eqref{eq_PsiConservation}} obtained by averaging over a finite number $S$ of independent numerical simulations with $N$ particles each.
If $N$ does indeed correspond to the actual number of particles in the system, the limit $S\to\infty$ converges to the true solution to {the} probabilistic formulation of the N-body problem, which is different from the collisionless Boltzmann equation~\eqref{eq_LiouvilleTheorem}.
The main differences with {respect to the traditional, deterministic interpretation of N-body simulations} are:
\begin{enumerate}
 \item {The probabilistic} interpretation is entirely based on the one-point probability distribution $\Psi$ (or, equivalently, the \emph{expected} phase-space density $\bar f$) instead of any particular \emph{individual} realisation where the phase-space density $f$ is given by a sum of Dirac deltas.
 \item  The evolution of individual discrete particles is not considered.
 \item $\Psi$ and $\bar f$ may be any distribution, including a perfectly smooth and continuous function, subject to the normalisation condition~\eqref{eq_PsiNormalization}: no coarse-graining procedure is involved, explicitly or implicitly, neither in their definition nor in equation~\eqref{eq_PsiConservation}.
 \item The acceleration in the collisionless Boltzmann equation~\eqref{eq_LiouvilleTheorem} can be deterministically computed from $f$, whereas in~\eqref{eq_PsiConservation} it is a random variable, whose statistical properties can be estimated from $\Psi$ and $N$. It is, therefore, a stochastic differential equation.
\end{enumerate}

\section{N-body simulations}
\label{sec_NBodySimulations}

\begin{figure*}
\centering
\begin{tabular}{c  c c c c c c}
& $t=0$ & $t=10$ & $t=100$ & $t=1000$ & $t=10^4$ & $t=10^5$ \\

\rotatebox{90}{\hspace{.7cm} $N=10$} &
\includegraphics[width=.135\textwidth]{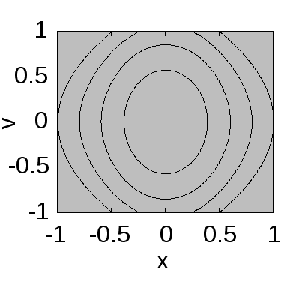} &
\includegraphics[width=.135\textwidth]{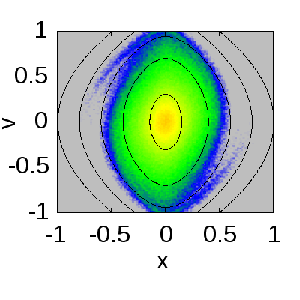} &
\includegraphics[width=.135\textwidth]{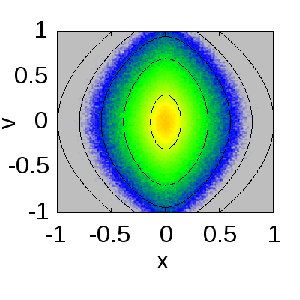} &
\includegraphics[width=.135\textwidth]{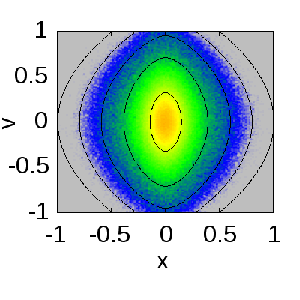} &
\includegraphics[width=.135\textwidth]{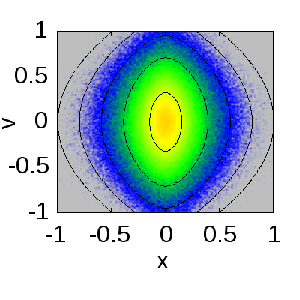} &
\includegraphics[width=.135\textwidth]{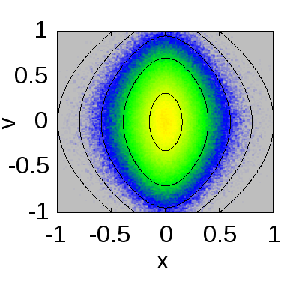}\\

\rotatebox{90}{\hspace{.7cm} $N=100$} &
\includegraphics[width=.135\textwidth]{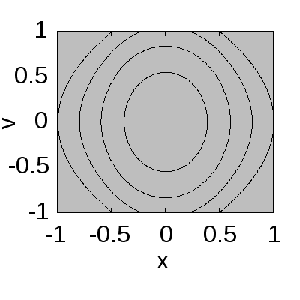} &
\includegraphics[width=.135\textwidth]{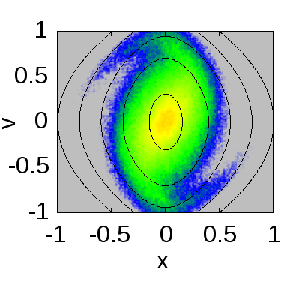} &
\includegraphics[width=.135\textwidth]{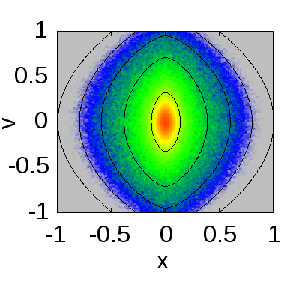} &
\includegraphics[width=.135\textwidth]{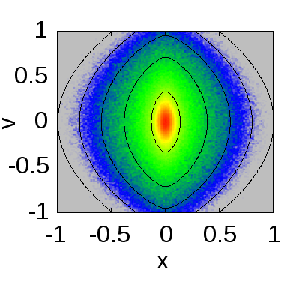} &
\includegraphics[width=.135\textwidth]{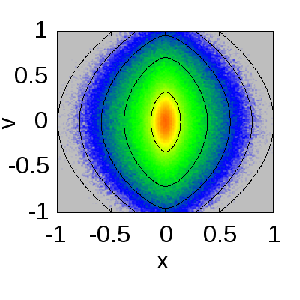}
&
\\

\rotatebox{90}{\hspace{.7cm} $N=1000$} &
\includegraphics[width=.135\textwidth]{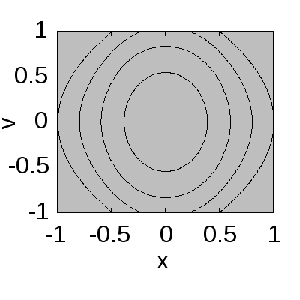} &
\includegraphics[width=.135\textwidth]{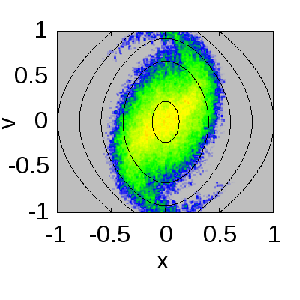} &
\includegraphics[width=.135\textwidth]{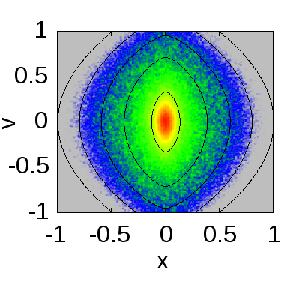} &
\includegraphics[width=.135\textwidth]{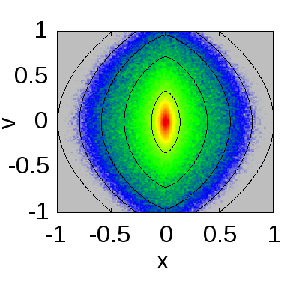} &
&
\\

& $t=0$ & $t=10$ & $t=100$ & $t=1000$ & $t=10^4$ & $t=10^5$ \\

\rotatebox{90}{\hspace{.7cm} $N=10$} &
\includegraphics[width=.135\textwidth]{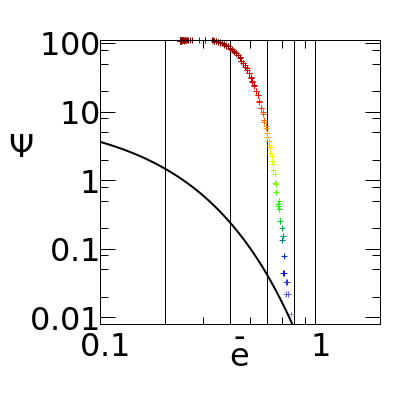} &
\includegraphics[width=.135\textwidth]{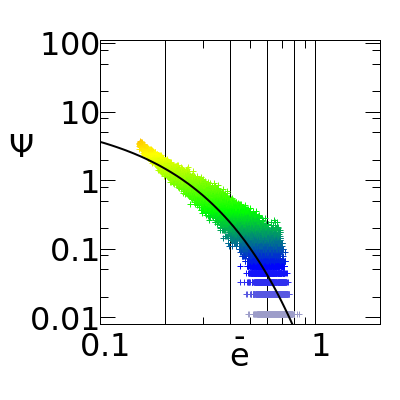} &
\includegraphics[width=.135\textwidth]{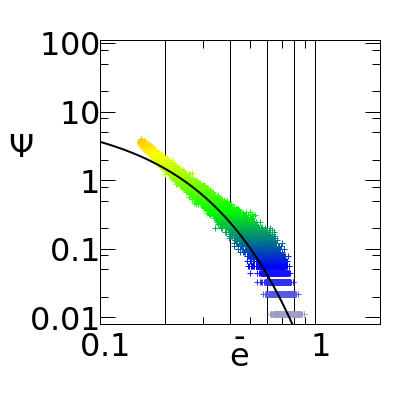} &
\includegraphics[width=.135\textwidth]{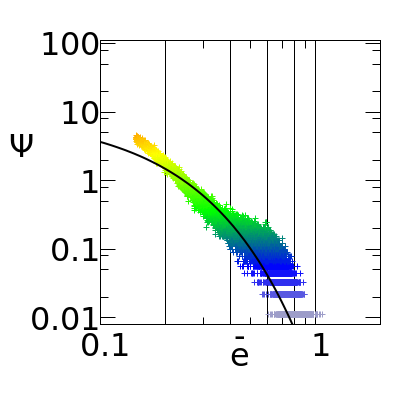} &
\includegraphics[width=.135\textwidth]{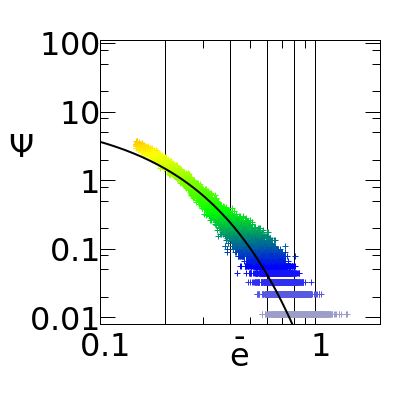} &
\includegraphics[width=.135\textwidth]{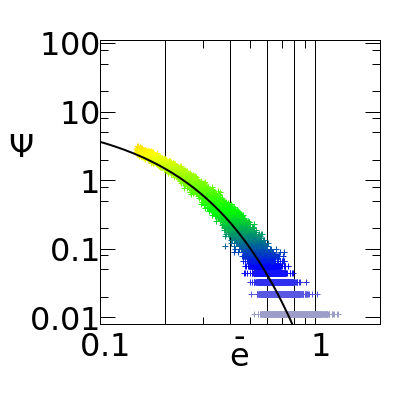}\\

\rotatebox{90}{\hspace{.7cm} $N=100$} &
\includegraphics[width=.135\textwidth]{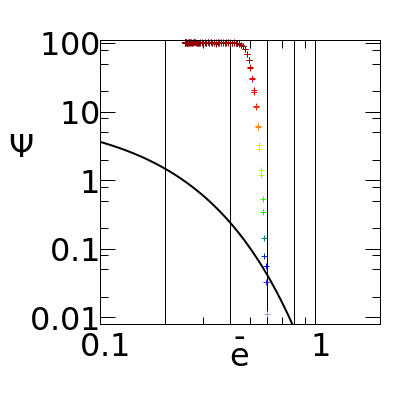} &
\includegraphics[width=.135\textwidth]{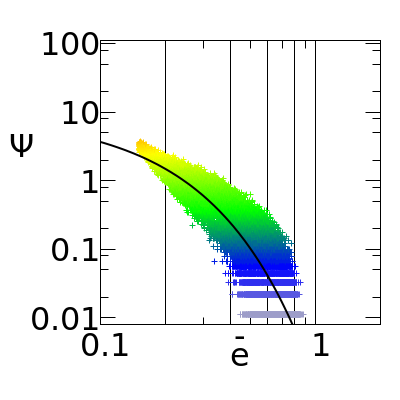} &
\includegraphics[width=.135\textwidth]{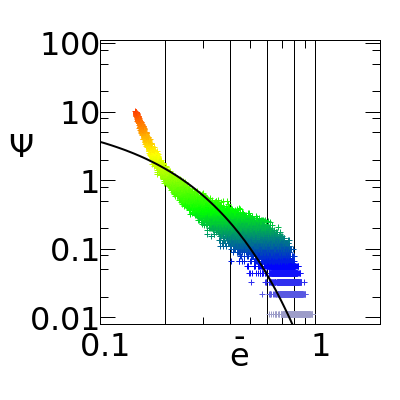} &
\includegraphics[width=.135\textwidth]{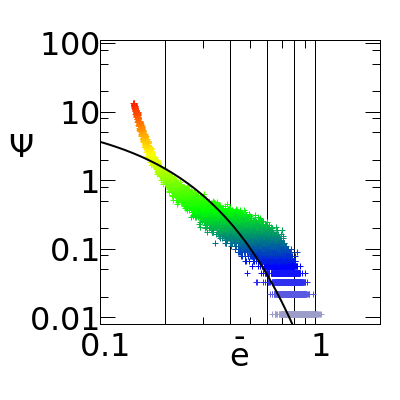} &
\includegraphics[width=.135\textwidth]{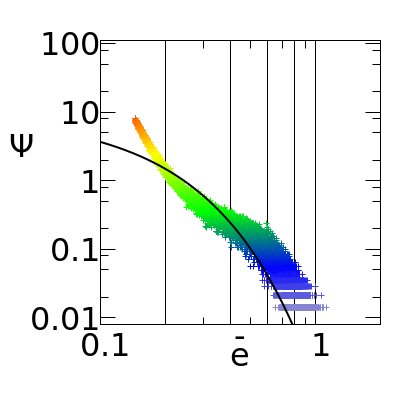}
&
\\

\rotatebox{90}{\hspace{.7cm} $N=1000$} &
\includegraphics[width=.135\textwidth]{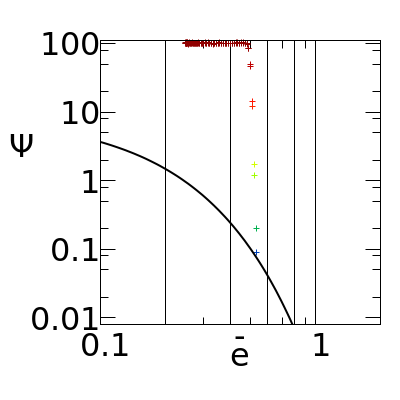} &
\includegraphics[width=.135\textwidth]{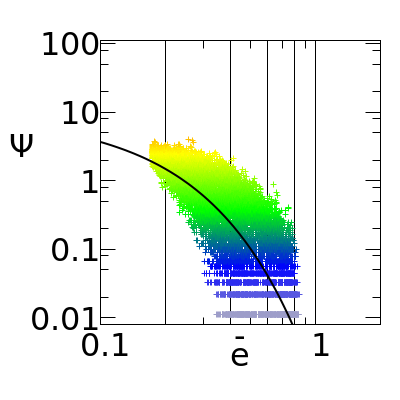} &
\includegraphics[width=.135\textwidth]{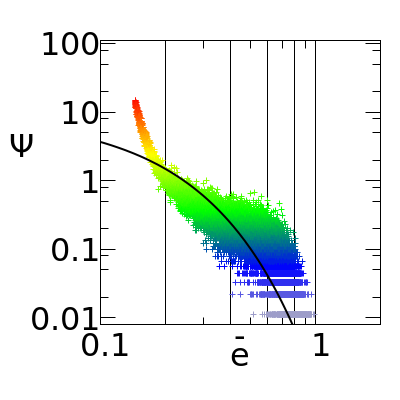} &
\includegraphics[width=.135\textwidth]{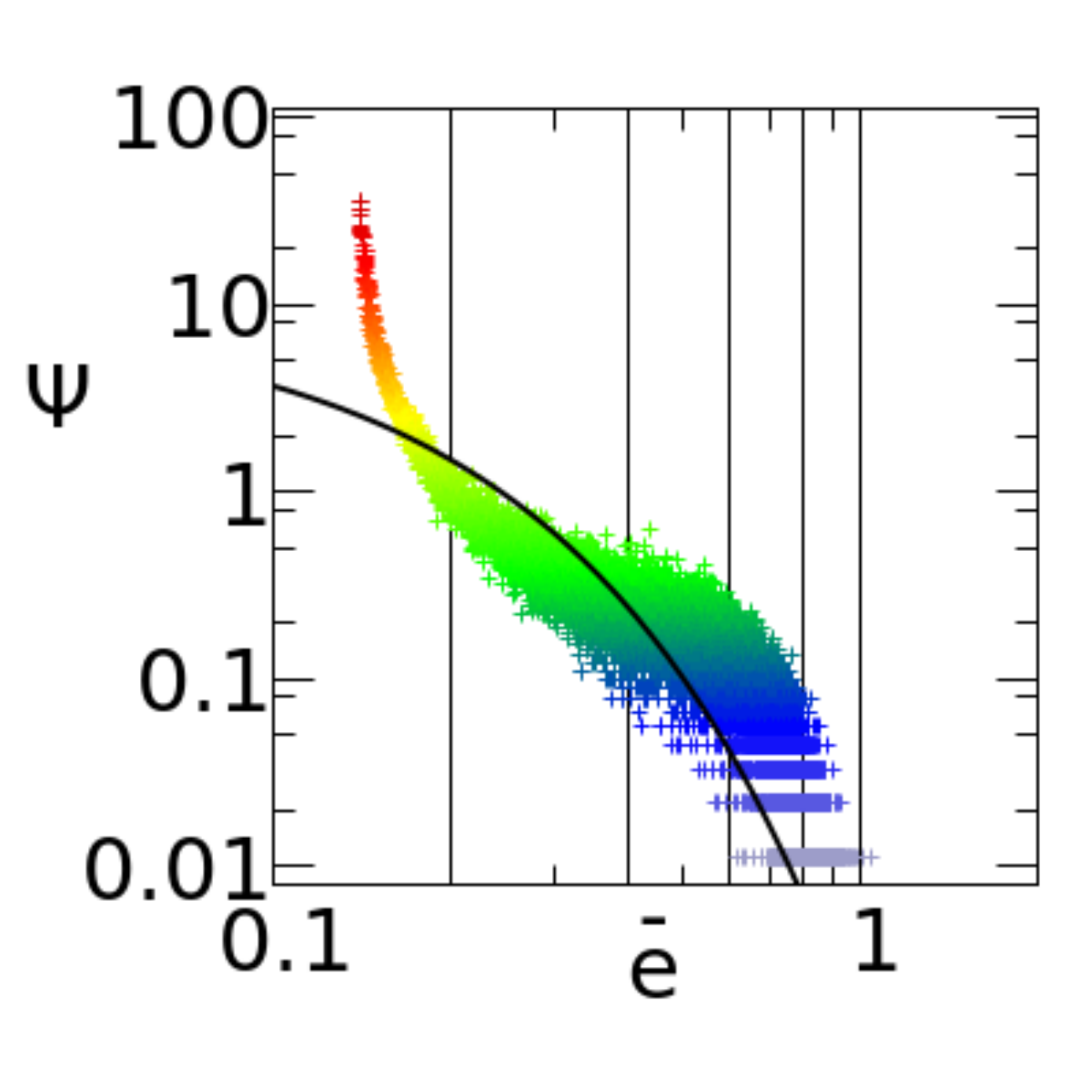} &
&
\\
\end{tabular}
\caption
{
Top panels: evolution of the probability density $\Psi(x,v,t)$ in the \emph{bar} case~\eqref{eq_barIC_Psi} for different times (columns) and number of particles (rows).
Colours represent probability density (varying by about four orders of magnitude), and contours have been drawn at energies $\e(x,v) = \frac{v^2}{2} + \bar{\phi}(x) = \{\, 0.2,\, 0.4,\, 0.6,\, 0.8,\, 1.0\, \}$.
Bottom panels: same as above, in terms of $\Psi(\e)$, using identical colour scale and vertical lines to indicate energy contour levels.
{The Maxwell-Boltzmann solution is shown as a black solid line for illustrative purposes.}
}
\label{Fig_Psi_bar}
\end{figure*}
\begin{figure*}
\centering
\begin{tabular}{c  c c c c c c}
& $t=0$ & $t=10$ & $t=100$ & $t=1000$ & $t=10^4$ & $t=10^5$ \\

\rotatebox{90}{\hspace{.7cm} $N=10$} &
\includegraphics[width=.135\textwidth]{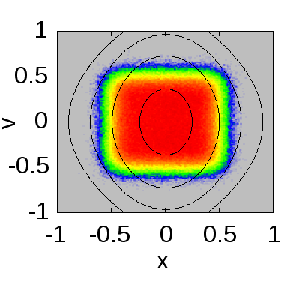} &
\includegraphics[width=.135\textwidth]{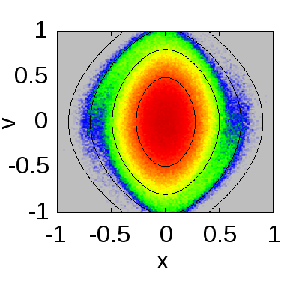} &
\includegraphics[width=.135\textwidth]{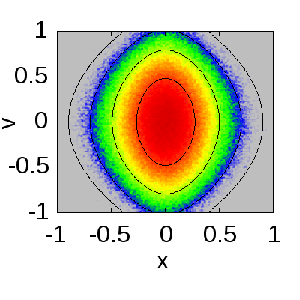} &
\includegraphics[width=.135\textwidth]{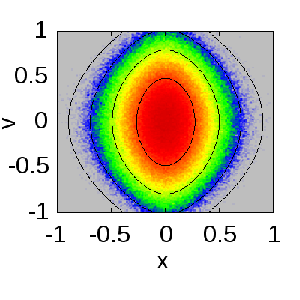} &
\includegraphics[width=.135\textwidth]{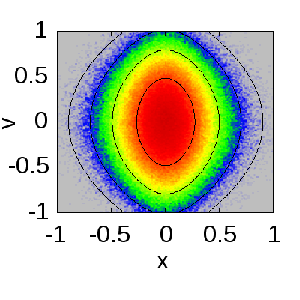} &
\includegraphics[width=.135\textwidth]{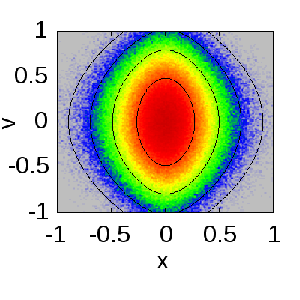}\\

\rotatebox{90}{\hspace{.7cm} $N=100$} &
\includegraphics[width=.135\textwidth]{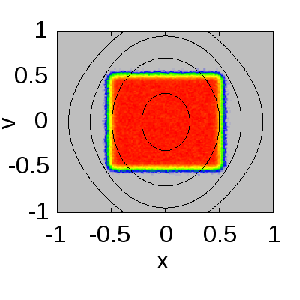} &
\includegraphics[width=.135\textwidth]{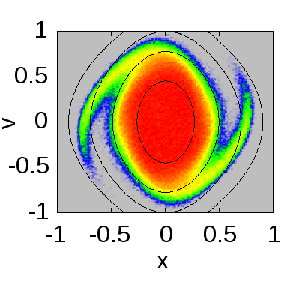} &
\includegraphics[width=.135\textwidth]{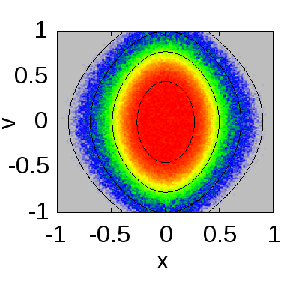} &
\includegraphics[width=.135\textwidth]{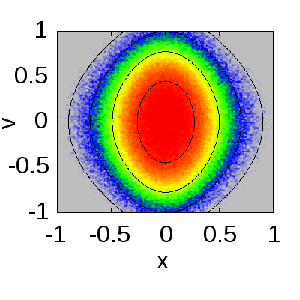} &
\includegraphics[width=.135\textwidth]{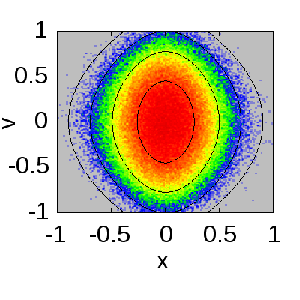}
&
\\

\rotatebox{90}{\hspace{.7cm} $N=1000$} &
\includegraphics[width=.135\textwidth]{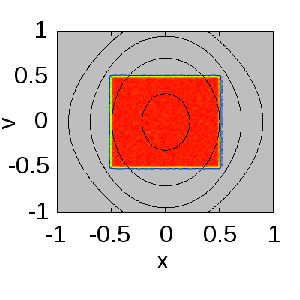} &
\includegraphics[width=.135\textwidth]{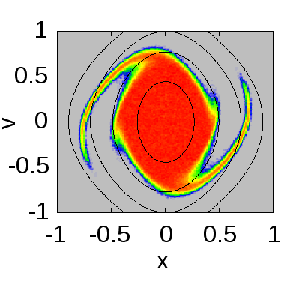} &
\includegraphics[width=.135\textwidth]{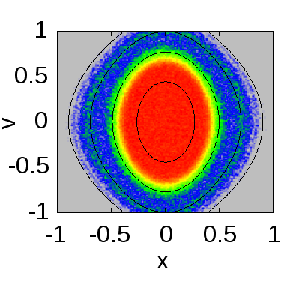} &
\includegraphics[width=.135\textwidth]{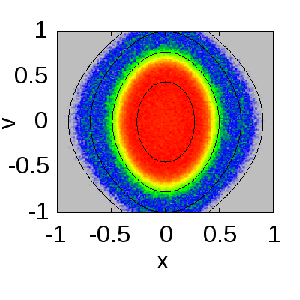} &
&
\\

& $t=0$ & $t=10$ & $t=100$ & $t=1000$ & $t=10^4$ & $t=10^5$ \\

\rotatebox{90}{\hspace{.7cm} $N=10$} &
\includegraphics[width=.135\textwidth]{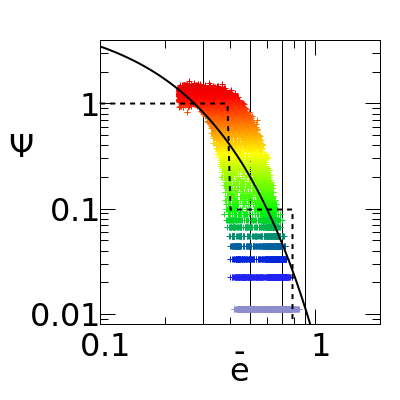} &
\includegraphics[width=.135\textwidth]{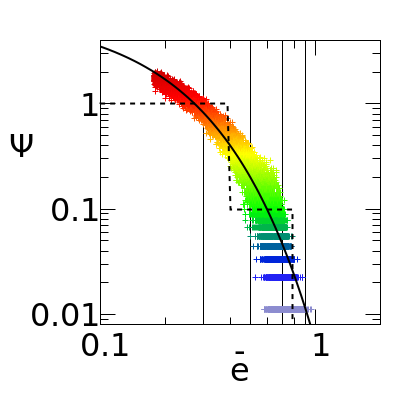} &
\includegraphics[width=.135\textwidth]{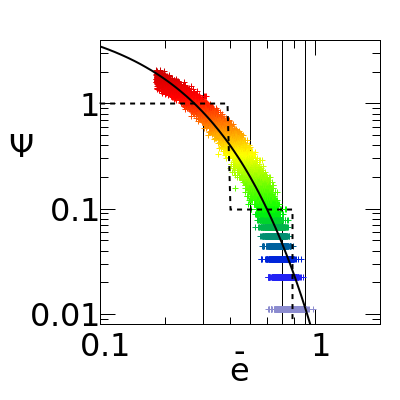} &
\includegraphics[width=.135\textwidth]{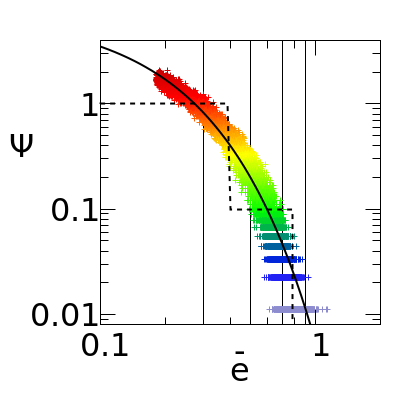} &
\includegraphics[width=.135\textwidth]{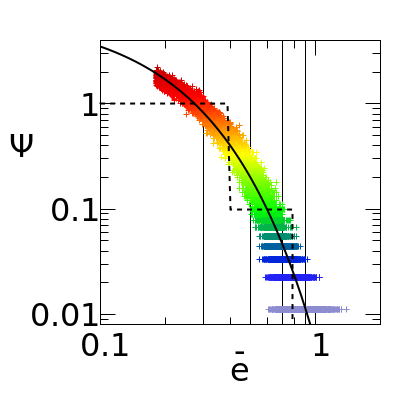} &
\includegraphics[width=.135\textwidth]{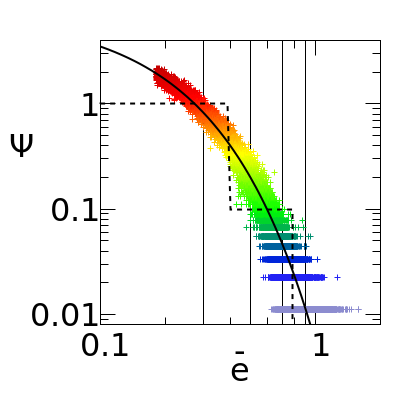}\\

\rotatebox{90}{\hspace{.7cm} $N=100$} &
\includegraphics[width=.135\textwidth]{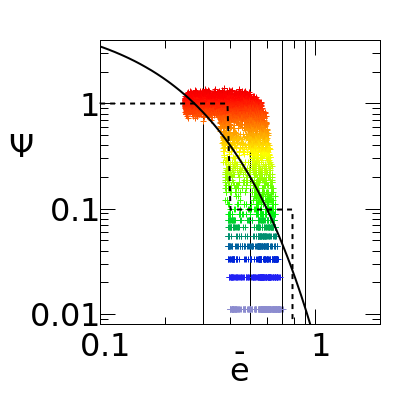} &
\includegraphics[width=.135\textwidth]{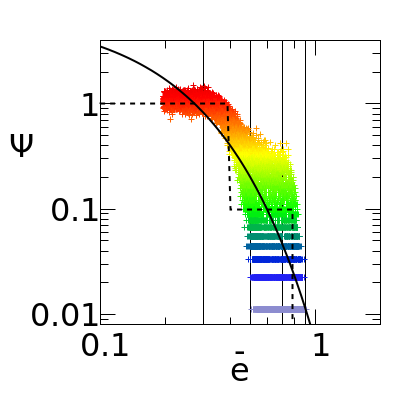} &
\includegraphics[width=.135\textwidth]{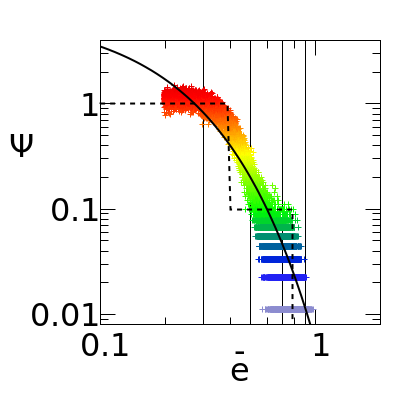} &
\includegraphics[width=.135\textwidth]{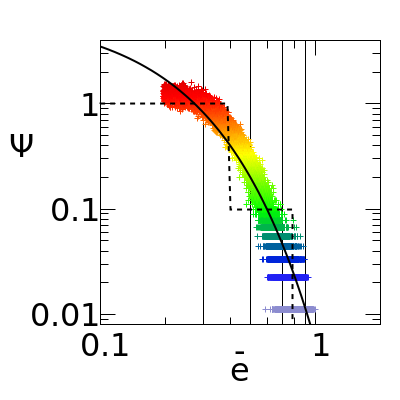} &
\includegraphics[width=.135\textwidth]{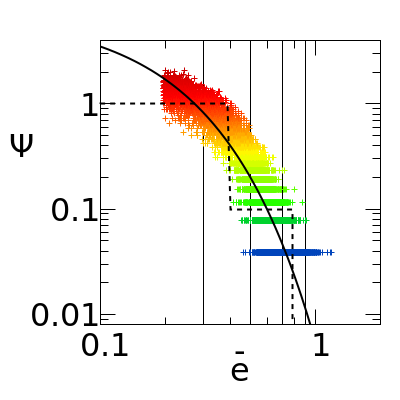} 
&
\\

\rotatebox{90}{\hspace{.7cm} $N=1000$} &
\includegraphics[width=.135\textwidth]{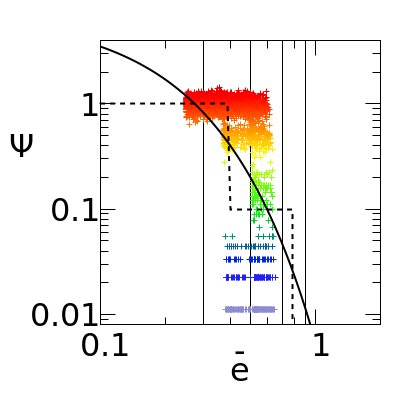} &
\includegraphics[width=.135\textwidth]{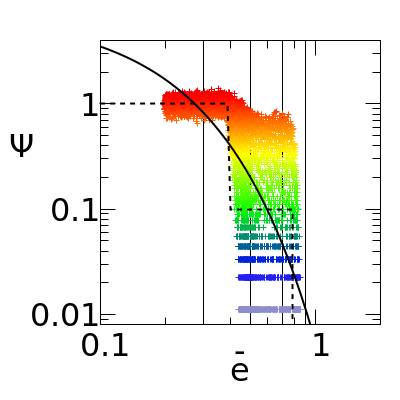} &
\includegraphics[width=.135\textwidth]{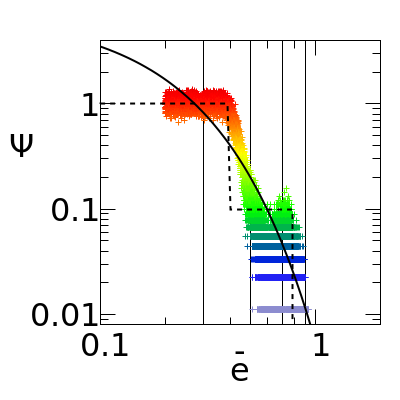} &
\includegraphics[width=.135\textwidth]{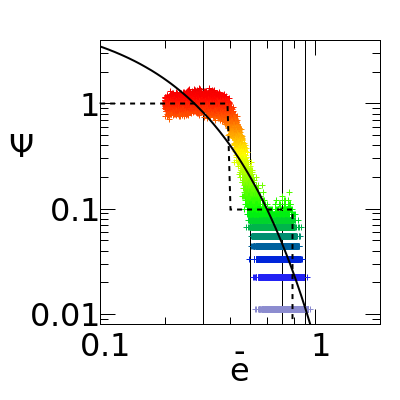} &
&
\\
\end{tabular}
\caption
{
Same as Figure~\ref{Fig_Psi_bar}, for the \emph{box} case~\eqref{eq_boxIC}.
Probability density (colours) span now two orders of magnitude and energy contours have been drawn at $\e = \{\, 0.3,\, 0.5,\, 0.7,\, 0.9\, \}$.
{In addition to the Maxwell-Boltzmann distribution, the ansatz proposed by~\citet{Teles+11} is also shown as a dashed line in the bottom panels.}
}
\label{Fig_Psi_box}
\end{figure*}

In order to carry out the $S$ independent simulations to be averaged, we have developed a code that solves the N-body problem in one dimension to very high accuracy.
More specifically, the trajectories of the particles are followed analytically between collisions, without any approximation, and numerical integration errors are limited to machine round-off precision \citep[cf.][and references therein]{Schulz+13}.

We consider two different initial conditions, both with the same distribution of positions (uniform random numbers between $-\frac{1}{2}$ and $\frac{1}{2}$) but different velocity distributions.
In the first initial condition, analogous to the periodic, clockwise-rotating \emph{bar} case~\eqref{eq_barIC_f} discussed in Section~\ref{ssec_PeriodicSolution}, velocities are set to zero
\begin{equation}
 \Psi(x,v,0) = H(x+\frac{1}{2})\ H(\frac{1}{2}-x)\ \delta(v),
 \label{eq_barIC_Psi}
\end{equation}
whereas the other initial condition has uniform random velocities between $-\frac{1}{2}$ and $\frac{1}{2}$.
In other words, the initial probability density resembles a \emph{box} in phase space
\begin{equation}
 \Psi(x,v,0) = H(x+\frac{1}{2})\ H(\frac{1}{2}-x)\ H(v+\frac{1}{2})\ H(\frac{1}{2}-v),
\label{eq_boxIC}
\end{equation}
and therefore we will refer to~\eqref{eq_barIC_Psi} and~\eqref{eq_boxIC} as the \emph{bar} and the \emph{box} cases, respectively.

For each initial condition, we have run sets of $S=\{\, {90000,}\, 9000,\, 900\, \}$ simulations with $N=\{\, {10,}\, 100,\, 1000\, \}$ particles each, so that the product $N S = 9 \times 10^5$ in all cases.
With the purpose of studying the long-term evolution of the system, we consider times up to $t=1000$, much larger than the characteristic time $\tcross = 2$ or the relaxation time $\trelax$ estimated from~\eqref{eq_RelaxTime3D} {for the three-dimensional case.
In one dimension, previous results suggest that our systems should have attained a quasi-stationary state by that time, but not thermal equilibrium \citep{JW10}, and we have let our simulations with $N=10$ and $N=100$ particles run up to $t=10^5$ and $t=10^4$, respectively, to explore the asymptotic approach to such limit}.
In order to study the dependence of the results with the number of particles over a broader range in $N$, we have also run an additional set of $S=50$ simulations of the \emph{bar} case with $N=10^6$ up to $t=10$.

For each individual simulation, we subtract the position and velocity of the centre of mass to place the system at the origin of phase space.
To measure $\Psi(x,v,t)$, we discretise the phase space in homogeneous subdivisions with $\Delta x= 0.01$ and $\Delta v=0.01$ and evaluate the probability density at fixed values of $x_i$ and $v_j$, given a set of times $t_k$.
For a single simulation $s$, the individual probability density
\begin{equation}
\psi^{(s)} (x_i,v_j,t_k) \equiv \psi^{(s)}_{ijk} = \frac{ n^{(s)}_{ijk} }{ N\, \Delta x\, \Delta v }
\label{eq_IndividualPsiDef}
\end{equation}
is proportional to the number of particles $n^{(s)}_{ijk}$ found inside the bin $(x_i \pm \frac{\Delta x}{2}, v_j \pm \frac{\Delta v}{2})$ at time $t_k$.
Formally, the probability density $\Psi(x,v,t)$ is given by the weighted average of all possible realisations of the initial condition.
Approximating the result as an average over a finite number of simulations
\begin{equation}
\Psi_{ijk} = \frac{1}{S} \sum_{s=1}^{S} \psi^{(s)}_{ijk}
\label{eq_MeanPsiDef}
\end{equation}
is simply a Monte Carlo method to evaluate such integral numerically.
Once this quantity has been obtained, we compute the mean velocity distribution
\begin{equation}
\bar{\eta}(v_j,t_k) \equiv \bar{\eta}_{jk} = \Delta x \sum_{i=1}^{N_x}\Psi_{ijk}
\label{eq_MeanVelDistr}
\end{equation}
and the mean number density
\begin{equation}
\bar{\rho}(x_i,t_k) \equiv \bar{\rho}_{ik} = \Delta v \sum_{j=1}^{N_v}\Psi_{ijk},
\label{eq_MeanDensity}
\end{equation}
where $N_x$ and $N_v$ correspond to the number of subdivisions in position and velocity, respectively.

\section{Results}
\label{sec_Results}

\begin{figure*}
\begin{tabular}{c c c}
& \LARGE{$\rho_{bar}(x)$} & \LARGE{$\rho_{box}(x)$} \\

\rotatebox{90}{\Large\hspace{2.7 cm}$N=10$} &
\includegraphics[width=.375\textwidth]{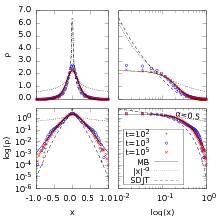} &
\includegraphics[width=.375\textwidth]{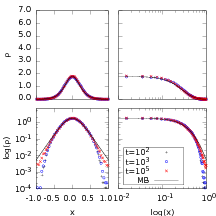} \\

\rotatebox{90}{\Large\hspace{2.7 cm}$N=100$} &
\includegraphics[width=.375\textwidth]{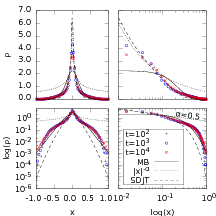} &
\includegraphics[width=.375\textwidth]{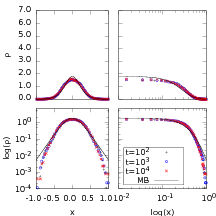} \\

\rotatebox{90}{\Large\hspace{2.8 cm}$N=1000$} &
\includegraphics[width=.375\textwidth]{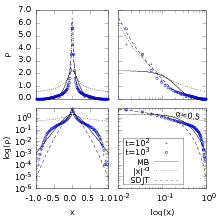} &
\includegraphics[width=.375\textwidth]{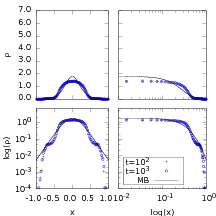}
\end{tabular}
\caption{
Dependence of the mean density profile $\bar{\rho}(x)$ on time and $N$, for both the \emph{bar} (left column) and \emph{box} (right column) cases{, compared to the Maxwell-Boltzmann solution (solid line). A {pure} power law $\bar{\rho}(x)\propto x^{-0.5}$ {(dotted line)} and the {power law + exponential form} proposed by~\citet{Schulz+13} (their equation 17{, dashed line}) have also been plotted for the \emph{bar} initial condition.}
}
\label{Fig_Rho}
\end{figure*}
\begin{figure*}
\begin{tabular}{c c c}
& \LARGE{$\eta_{bar}(v)$} & \LARGE{$\eta_{box}(v)$} \\

\rotatebox{90}{\Large\hspace{2.7 cm}$N=10$} &
\includegraphics[width=.375\textwidth]{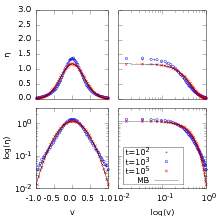} &
\includegraphics[width=.375\textwidth]{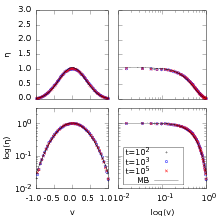} \\

\rotatebox{90}{\Large\hspace{2.7 cm}$N=100$} &
\includegraphics[width=.375\textwidth]{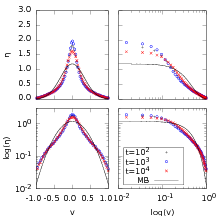} &
\includegraphics[width=.375\textwidth]{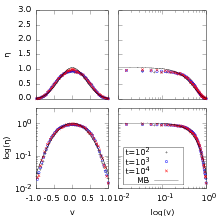} \\

\rotatebox{90}{\Large\hspace{2.8 cm}$N=1000$} &
\includegraphics[width=.375\textwidth]{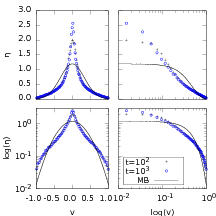} &
\includegraphics[width=.375\textwidth]{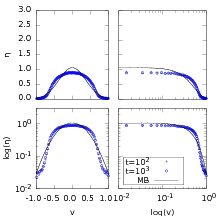}
\end{tabular}
\caption{
Dependence of the mean velocity distribution $\bar{\eta}(x)$ on $t$ and $N$.
Colours and line styles are identical to those in Figure~\ref{Fig_Rho}.
}
\label{Fig_Eta}
\end{figure*}

Figures~\ref{Fig_Psi_bar} and~\ref{Fig_Psi_box} show the evolution of the probability density $\Psi$ at times $t=\{\,0,\,10,\,100,\,1000,\,{10^4,\, 10^5\,}\}$ for $N=\{\,{10},\, 100,\, 1000\, \}$ in the \emph{bar} and \emph{box} cases, respectively.
Top panels represent it as a function of the phase-space coordinates $(x,v)$, whereas bottom panels illustrate the dependence on the expected energy per unit of mass
\begin{equation}
{\e(x,v) = \frac{v^2}{2} + \bar{\phi}(x)}.
\end{equation}
The mean density $\bar{\rho}(x)$ and the velocity distribution $\bar{\eta}(v)$ are plotted in Figures~\ref{Fig_Rho} and~\ref{Fig_Eta}, respectively, for {$t=100$ and $1000$, as well as $t=10^4$ (for $N=100$) and $10^5$ (for $N=10$).}

As anticipated in Section~\ref{sec_ProbabilityDensity}, the evolution of $\Psi$ bears clear resemblances with the predictions of the collisionless Boltzmann equation, but also important differences.
Our results clearly show that the shape of the initial configuration is not only distorted, but also that $\Psi$ is not conserved throughout the evolution of the system.
In a deterministic fluid, whose dynamics is described by the collisionless Boltzmann equation, the distribution function $f$ \emph{must} be strictly conserved along Lagrangian trajectories, and any departure from such behaviour can only be a consequence of coarse graining and/or numerical errors.
{In contrast}, diffusion in phase space of the one-point probability $\Psi$ is a \emph{physical} effect, and it is completely unrelated to the coarse-graining that has been performed in order to produce the estimates that are represented in all our figures.
Furthermore, the central probability density \emph{increases} in the \emph{box} case as the system evolves towards the Maxwell-Boltzmann distribution (see Figure~\ref{Fig_Psi_box}), which implies the action of additional terms/processes other than diffusion {\citep[see e.g.][]{Chavanis13}}.

After a {relatively short time, of the order of $\sim 10-100$ crossing times}, the combined effects of distortion (phase mixing) and (stochastic) diffusion seem to drive the probability density towards a quasi-stationary state that admits a description of the form $\Psi(\e)$ {as the system slowly evolves towards thermal equilibrium}.
This qualitative behaviour is common to all the cases we have considered.
However, there are crucial differences between the two initial conditions, as well as a certain dependence on the number of particles $N$.

The results obtained for our \emph{bar} and \emph{box} cases demonstrate that the shape of the function $\Psi(\e)$ {that describes the quasi-stationary state} is certainly \emph{not} universal, and it clearly retains memory of the initial conditions {for a much longer time than commonly believed (basically, as long as it is appreciably different from the thermodynamic equilibrium limit)}.
Not only the shapes of $\Psi(\e)$ on the bottom panels of Figures~\ref{Fig_Psi_bar} and~\ref{Fig_Psi_box} are markedly different; the values of $\Psi(x,v)$ on the top panels, as well as the density and velocity distributions plotted in Figures~\ref{Fig_Rho} and~\ref{Fig_Eta}, show that the probability density of the {quasi-stationary} state in the \emph{bar} case is much more concentrated near the origin of phase space (particles close to the centre, moving at slow velocities) than in the \emph{box} case (see the discussion in the next Section).
{On very long time scales, the quasi-stationary probability density seems to evolve asymptotically towards the Boltzmann distribution in both cases}.

{In addition, we also detect systematic effects on the shape of the probability density $\Psi(\e)$ during the quasi-steady phase associated to the number of particles}.
First and foremost, the dispersion in phase space seems to be slightly faster for low values of $N$, qualitatively consistent with the idea that the amplitude of random fluctuations in the acceleration~\eqref{eq_TrueAcc} scales as $\frac{1}{\sqrt{N}}$.
This can be seen in the early evolution of $\Psi(x,v,t)$ (the thickness of the distortions {at $t=10$} on the top panels of Figures~\ref{Fig_Psi_bar} and~\ref{Fig_Psi_box}) as well as on the time it takes the system to approach an apparently steady state.
{We find, in agreement with previous results \citep[e.g.][]{JW10}, that violent relaxation to the quasi-stationary state occurs in a few tenths of dynamical times, orders of magnitude faster than the evolution towards thermal equilibrium, but our results show a small, yet significant, dependence on $N$}.
We do not observe any relevant change in the density or the velocity distribution of the \emph{box} case {between $t=100$ and $t=1000$} for any value of $N$, although the phase-space structures (evident as tails on the top panels of Figure~\ref{Fig_Psi_box} and bumps at $\e\sim 0.8$ on the bottom panels, {as well as a systematic change in the shape of $\bar{\rho}(x)$ and $\bar{\eta}(v)$ in Figures~\ref{Fig_Rho} and~\ref{Fig_Eta}}) survive for a longer time as the number of particles is increased.
On the other hand, the \emph{bar} shows {a slightly more complex evolution, where the maximum values of $\Psi(\e)$, $\bar\rho(x)$, and $\bar\eta(v)$ increase (the particle distribution becomes more concentrated) up to a certain time (that increases with $N$), and then the maxima decrease again as the system approaches the thermal solution.}

Besides {setting the characteristic times}, the value of $N$ also affects the details of the quasi-stationary state.
On the one hand, the maxima of $\Psi(\e)$, $\bar\rho(x)$, and $\bar\eta(v)$ decrease with the number of particles for the \emph{bar} case.
On the other hand, the \emph{box} shows subtle differences in the opposite direction: for {$N=10$ and $100$}, the maximum of $\Psi$ is located at the origin of positions and velocities, but for {larger} values of $N$, it occurs along a ring in phase space, roughly corresponding to $\e\sim 0.3$ (bottom panel of Figure~\ref{Fig_Psi_box}).
This translates into $\bar{\rho}(x)$ and $\bar{\eta}(v)$ being slightly {shallower in the central parts for larger $N$} in Figures~\ref{Fig_Rho} and~\ref{Fig_Eta}.

\begin{figure*}
	\centering
	\begin{tabular}{c c c}
			& {\Large\textbf{t=6}} & {\Large\textbf{t=10}} \\
			
	\rotatebox{90}{\hspace{3.cm} \Large{$N=10$}} &
	\includegraphics[width=.4\textwidth]{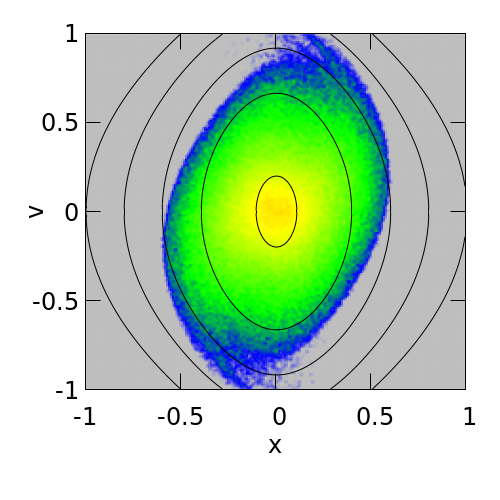} &
	\includegraphics[width=.4\textwidth]{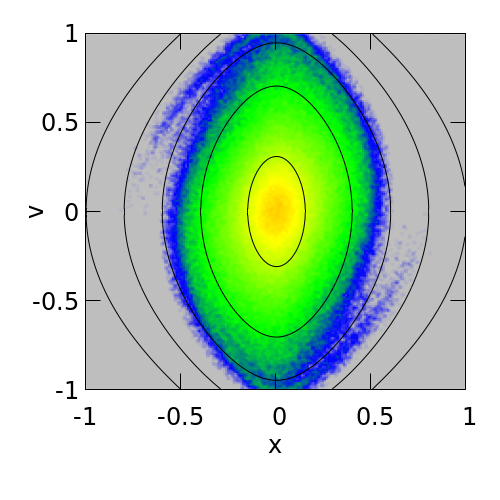} \\

	\rotatebox{90}{\hspace{3.cm} \Large{$N=1000$}} &
	\includegraphics[width=.4\textwidth]{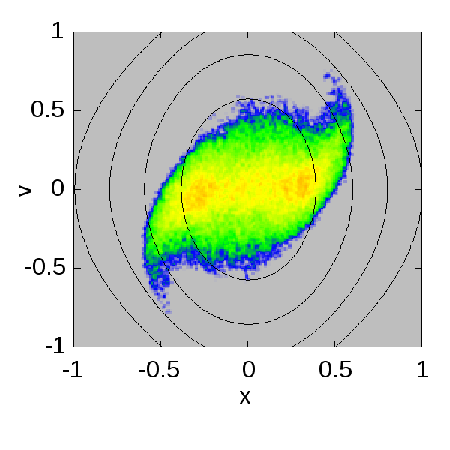} &
	\includegraphics[width=.4\textwidth]{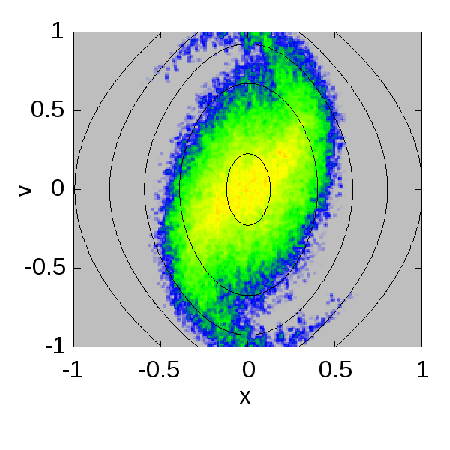} \\
	
	\rotatebox{90}{\hspace{3.cm} \Large{$N=10^6$}} &
	\includegraphics[width=.4\textwidth]{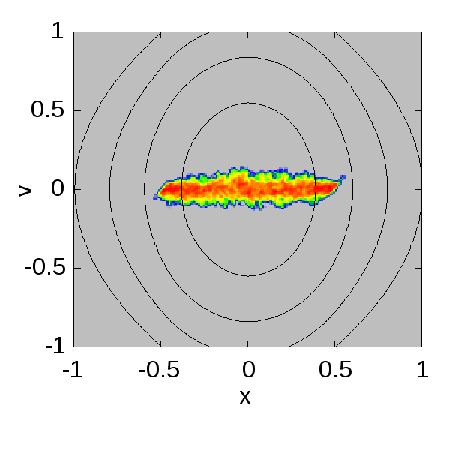} &
	\includegraphics[width=.4\textwidth]{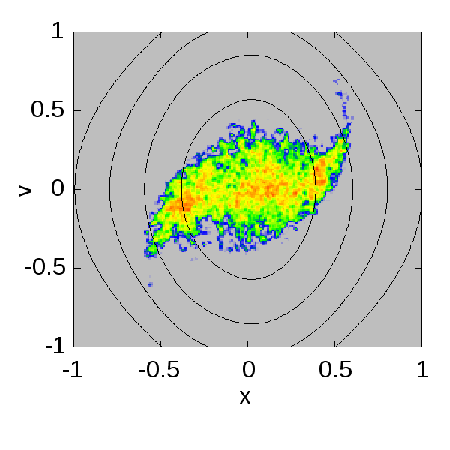}
	\end{tabular}
	\caption
	{$\Psi(x,v,t)$ in the \emph{bar} case for $t = \{\, 6,\, 10\, \}$ and  $N = \{\, {10},\, 1000,\, 10^6\, \}$.
	Colour scale and contour levels are identical to those in Figure~\ref{Fig_Psi_bar}.
	}\label{Fig_million}
\end{figure*}
\begin{figure*}
	\centering
	\begin{tabular}{c c c}
	\includegraphics[width=.3\textwidth]{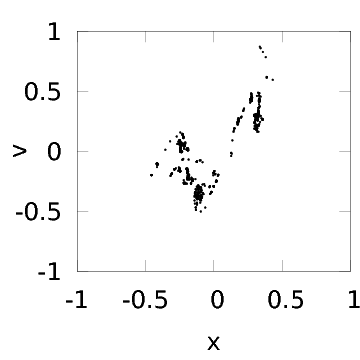} &
	\includegraphics[width=.3\textwidth]{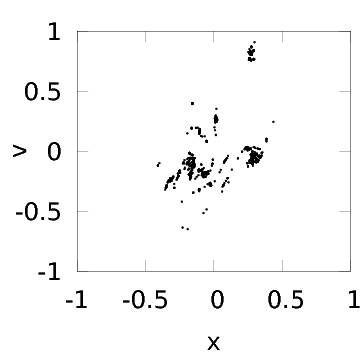} &
	\includegraphics[width=.3\textwidth]{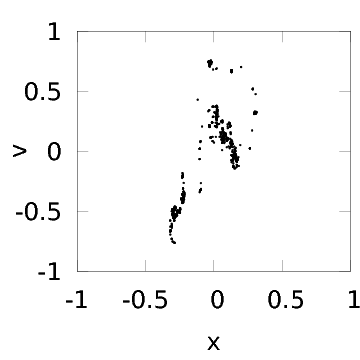} 
	\end{tabular}
	\caption
	{{Snapshots of three different individual simulations of the \emph{bar} case with $N=1000$ particles at $t=10$.}
	}\label{Fig_individual}
\end{figure*}

In order to address the dependence with $N$ over a broader range and shed some light on the possible extrapolation towards $N\to\infty$, we show in Figure~\ref{Fig_million} the probability density $\Psi(x,v,t)$ for values of $N$ up to $10^6$ (representative, for instance, of the number of stars in a super-massive star cluster or a tiny dwarf galaxy).
These data make it clear that even the small differences observed between {10}, 100, and 1000 particles are indeed significant, with the most relevant aspect being the slower diffusion in phase space.

It is also noteworthy that {friction and diffusion slow} down the rotation of the \emph{bar} in phase space.
At $t=6$, only the $N=10^6$ simulations have reached the maximum expansion state (a horizontal bar), whereas the smaller-N simulations display a thicker configuration at a certain angle with respect to the $x$ axis.
At $t=10$, the system is almost at maximum contraction (a vertical \emph{bar}) for {$N=10$}, and there is some delay with respect to the periodic solution of the collisionless Boltzmann equation for $N=10^6$.
In fact, the results for $N=10^6$ at $t=10$ are fairly similar to the $N=1000$ counterpart at $t=6$ (i.e. one revolution less around the origin), {and a similar parallelism is observed between $N=1000$ and $N=10$.

These} results show that the characteristic times for both diffusion and phase lag with respect to the periodic solution of the collisionless Boltzmann equation scale very slowly with the number of particles.
Although we caution that logarithmic scaling is far from proven, Figure~\ref{Fig_million} suggests that {two or} three orders of magnitude in $N$ are roughly equivalent to a factor of about two in time.

\section{Discussion}
\label{sec_Discussion}

The problem of gravitational collapse in one spatial dimension has been widely studied in the literature under the {deterministic interpretation, often focusing on the limit $N\to\infty$}.
Particular analytical solutions of the collisionless Boltzmann equation for the distribution function $f(x,v,t)$ have been derived under different assumptions \citep[e.g.][]{Fillmore&Goldreich84, Bertschinger85, Alard13, Colombi15}, and several attempts have been made to solve it numerically without resorting to the N-body approximation \citep[e.g.][]{Hahn+13, Yoshikawa+13, Colombi&Touma14, Hahn&Angulo16, Mocz&Succi17, Sousbie&Colombi16}.
{For finite $N$}, initial conditions similar to our \emph{bar} and \emph{box} cases have previously been considered, and the reported results are fairly consistent with our findings if we identify the distribution function with the expected average $\bar f(x,v,t)$, weighted over all possible realisations of the \emph{random} initial conditions.
{Here we will discuss some of the main parallelisms and differences between our numerical results and those obtained in the deterministic formulation}.

\subsection{{Averaging}}

Averaging {over a finite number $S$ of independent simulations is the} key concept {that we would like to advocate for}.
From a theoretical point of view, it provides a well-defined and physically-motivated prescription to go from a fine-grained, discrete description of the system, where the distribution function $f(x,v,t)$ is a sum of Dirac deltas, to a probabilistic description in terms of the probability density $\Psi(x,v,t)$ or, equivalently, the \emph{expected} phase-space density $\bar f(x,v,t)$, that is most likely continuous and smooth in any realistic system.

In practice, some authors \citep[e.g.][]{JW10, Schulz+13, Levin+14} do average over several simulations with slightly different number of equispaced particles and/or over a given time interval in order to reduce noise.
{The probabilistic interpretation of the N-body problem} provides not only a rigorous justification for a similar procedure, but, more importantly, ensures that the ensemble of simulations that are averaged are statistically independent and that the true solution is recovered in the limit $S\to\infty$.

When averaging over a short time interval, adjacent snapshots are certainly correlated; if the time interval is long compared to the dynamical time, correlations will decrease (albeit not vanish), but the evolution of the system may start to play a role.
Arguably, the correlation between simulations with slightly different values of $N$ should decay after a sufficiently long time, but a careful analysis should be carried out in order to determine the consequences of averaging over such an ensemble.
In both cases, it is clear that it is only possible to average over a limited number $S$ of simulations that cannot be made arbitrarily large.

In {contrast}, different realisations of the initial conditions will remain statistically independent at all times.
Since they are all evaluated at exactly the same precise time $t$, there is no downside to increasing the number of simulations, and the only limit to $S$ is imposed by the available CPU time.
In principle, one would expect that statistical fluctuations of $\bar f(x,v,t)$ around the true solution {for a given $N$} should decrease as $S^{-1/2}$.

{
On the other hand, another important aspect of the probabilistic formulation is that averaging over a finite ensemble of individual random realisations of the initial condition is merely a numerical Monte Carlo method of integrating the stochastic differential equation~\eqref{eq_PsiConservation} to compute the evolution of the continuous distribution $\Psi(x,v,t)$.
Contrary to the traditional paradigm, concerned with the precise trajectories of the $N$ particles that make up the system, here only the probability density of finding one of them after a time $t$, given the initial condition $\Psi(x,v,0)$, is tracked.
As a matter of fact, there are no actual particles -- let alone well-defined particle locations or trajectories -- in this interpretation of the problem.

This is a remarkable difference from the philosophical point of view that also brings important practical consequences.
In the deterministic approach, one can get extremely complex (even chaotic) behaviours that depend on the very finest details of the initial condition, and averaging over independent realisations provides only a partial view of what can actually happen in every one of them.
As shown in Figure~\ref{Fig_individual}, the distribution function $f(x,v)$ of each individual simulation in our ensemble is much more intricate than the smooth and continuous one-point probability density $\Psi(x,v)$.
This is, once again, unrelated to any coarse-graining $\hat f(x,v)$ over a suitably-defined small volume, which would yield a smooth, but still complex structure.
In other words, $f$, $\hat f$, and $\bar f$ are three different quantities, and their evolution is a different problem from both the mathematical and physical points of view.
}

{
In particular, small substructures (physical or numerical) in $f(x,v,0)$ survive in phase space (and often grow) for a very long time, and the complexity of the system (e.g. number of independent substructures) tends to increase with the number of particles. 
In contrast, the effects of Poisson noise go to zero as $S$ and/or $N$ approach infinity, even from the very initial instant $t=0$, whereas physical perturbations around a smooth $\Psi(x,v,0)$ diffuse away on a characteristic time scale that increases monotonically with particle number (see below).

Although it is likely that both approaches converge as $N\to\infty$, a careful definition of the problem under study is necessary for finite $N$, especially as far as substructure is concerned.
If one were interested, for instance, in the statistics of substructure in galactic halos, it is not obvious that $\Psi$ or $\bar f$ are the most relevant quantities.
Moreover, one should also ensure that all structures of interest in phase space are sufficiently well sampled.
Although the precise requirements in terms of $N$ and $S$ are difficult to quantify, it is clear that running many realisations with poor mass resolution will be sufficient to trace arbitrarily small structures in the initial conditions, but their survival time will be severely underestimated \citep[somewhat akin to the `overmerging problem' in traditional N-body simulations;][]{Klypin+99}.
}

\subsection{{Dependence with N}}
\label{ssec_Dependence_with_N}

From a computational point of view, it is much more advantageous to run additional simulations than to use a larger number of particles, both in terms of memory requirements (a simulation with small $N$ fits in any laptop computer) as well as CPU time.
Even if the N-body algorithm scaled as $\mathcal{O}(N)$ -- rather than e.g. $\mathcal{O}(N \log N)$ --, communication between different processors, if needed, would impose an overhead with respect to running $S$ independent simulations, which is an embarrassingly parallel problem.
Moreover, increasing $N$ yields a much more complex evolution due to the formation of a larger number of substructures on smaller and smaller scales.
In each individual simulation, phase space is better resolved, and the dynamics of the system is \emph{physically} more intricate.
The time between particle collisions near the centre of every single substructure becomes very short, and the number of time steps required to reach a given time $t$ {in every single realisation of the initial conditions increases much faster than linearly with $N$. Thus, the total CPU time to run $S$ simulations scales more steeply than $NS$.}

{According to the results presented in Section~\ref{sec_Results}, the evolution of one-point probability density $\Psi(x,v,t)$, once these individual details are averaged out, has a (to some extent, surprisingly) small dependence with $N$, and the} number of particles within any of our cells of size $\Delta x \Delta v$ in phase space\footnote{One may decrease noise and reach a lower measurable density $\bar\Psi_{\rm min} = \frac{1}{\Delta x \Delta v}$ by increasing the cell size, at the expense, of course, of a coarser graining.} is roughly proportional to the product $NS$ {over most of the region in phase space occupied by the system.
If the evolution of the probability density $\Psi$ were completely independent on the number of particles, increasing $N$ would be equivalent to increasing the number $S$ of simulations.
However, the standard deviation of the acceleration decreases as $N^{-1/2}$, and we find, consistently with all the previous studies carried out under the deterministic interpretation, that the evolution of $\Psi(x,v,t)$ does indeed depend on the number of particles, although the trends with $N$ are much less relevant, for instance, than the effect of the initial condition}.

Understanding the dependence with $N$ is not only an extremely interesting academic question, but it is also of the utmost importance in the context of N-body simulations where ${N_{\rm tracers}}\ll N$, both under the traditional and the probabilistic interpretations of the problem.
{There are, in our opinion, two possibilities:
\begin{enumerate}
 \item If the dependence with $N$ was, as we claim, weak, averaging over a large number $S$ of independent simulations may yield a more accurate estimation of $\Psi(x,v,t)$ than increasing $N$, at the expense of losing all information about individual particles (e.g. trajectories) and two-point or higher-order correlations (e.g. substructure).
 \item If it were strong, the results of any simulation with ${N_{\rm tracers}} \neq N$ (or the solution of the Vlasov-Poisson equation) would be completely misleading.
\end{enumerate}
}

{In any case}, we would like to argue that physical systems are characterised by a large -- yet \emph{finite} -- number of particles, {and therefore} investigating the dependence with $N$ {of the quasi-steady state over a broad range in $N$ may offer more insight than finding an accurate solution to the collisionless Boltzmann equation, valid {only} for a continuous fluid.}

\subsection{{Structure of the quasi-stationary state}}
\label{ssec_structure}

{In the {deterministic} interpretation, it has often been reported} that cold initial conditions -- all particles located along a thin line in phase space with negligible velocity dispersion -- {with a sinusoidal velocity perturbation (e.g. a peak of the primordial density field in the cosmological context)} yield density profiles that are well described by a power law $\rho(x)\propto x^{-\alpha}$ with logarithmic slope $\alpha \simeq 0.5$ \citep[e.g.][]{Binney04}, $\alpha \simeq 0.47$ \citep[e.g.][]{Schulz+13}, or $\alpha \simeq 0.4$ \citep[e.g.][]{Colombi&Touma14} over a {certain} radial range.

{Without introducing such fluctuation, one} would recover the periodic solution described in Section~\ref{ssec_PeriodicSolution}, valid both for a continuous fluid as well as for $N$ equispaced particles.
{Under} the {probabilistic} interpretation, {though}, any system with $N$ randomly distributed particles will slowly diffuse out from the initially cold configuration and end up in a quasi-stationary state of the form $\Psi(\e)$ where $\rho(x)$ is strongly peaked towards the centre, {even if all the particles start} with strictly null velocity and random uniform position between $x=-\frac{1}{2}$ and $x=\frac{1}{2}$.

{However, we do not think that a power law provides a suitable description of the density profile in our \emph{bar} case}.
One of the aspects that is most sensitive to the number of particles in the system is the presence of a `core' (where $\bar f$, $\bar\rho$, and $\bar\eta$ are constant) sufficiently close to the origin of phase space.
Analogous results have been reported in previous studies, based on both N-body simulations with equispaced initial conditions as well as other schemes to solve the Vlasov-Poisson equation {\citep[see e.g.][for a detailed discussion of the variation with distance of the logarithmic slope]{Colombi&Touma14}.}
The physical origin of this core is, however, still debated, and several effects that contribute to its appearance have been put forward:

As pointed out by \citet{Binney04}, discreteness prevents an accurate sampling of the `true' continuous distribution function, and the interplay between phase mixing and violent relaxation is `prematurely' terminated.
While this is certainly an artefact for ${N_{\rm tracers}}\ll N$, it represents a real physical process for the correct value of $N$, and we propose that determining the extent of the core as a function of particle number could actually provide an estimate of this quantity in astrophysical N-body systems, such as star clusters, galaxies, or even dark matter haloes.

On the other hand, velocity dispersion \citep[e.g.][]{Tremaine&Gunn79, Melott83, Teles+11, JW11, Colombi&Touma14} also leads to the development of a central core, often attributed to the fact that, according to Liouville's theorem, the fine-grained distribution function in phase space cannot increase, and therefore the finite initial value places a strong upper limit at any later time.
This is extremely relevant for our \emph{box} case, and \citet{Teles+11} proposed a theoretically-motivated \emph{ansatz} for the analytical form of the quasi-stationary $f(\epsilon)$ consisting on a central value $f_0$ up to some `Fermi level' $\epsilon_0$ delimiting the core and another constant value $f_{\rm h}<f_0$ up to the maximum energy $\epsilon_{\rm h}$ attained by the particles in the outer, diffuse halo.
This is broadly consistent with our numerical results, in the sense that the \emph{box} case presents a fairly well-defined core where $\bar f(\e) \sim 1$ (the initial value), but, rather than a sharp transition to an outer halo, Figure~\ref{Fig_Psi_box} suggest that the system evolves towards a state where $\bar f$ is a gradually decreasing function of $\e$ in the outer parts, reminiscent of the exponential cut-off predicted by Maxwell-Boltzmann or Lynden-Bell statistics \citep[see e.g.][]{Levin+14}, which is also observed in the \emph{bar} case \citep[cf.][]{Schulz+13}.

Finally, the growth of substructure from initially small fluctuations of a cold state induces a similar effect on the particle distribution in the central regions.
As shown by \citet{Schulz+13}, adding a short-wavelength perturbation of reduced amplitude to an infinitely thin sheet in phase space breaks the self-similarity of the collapse and the power-law density profile of the final state.
Similar results are reported by \citet{Colombi&Touma14}, where the extent of the core depends on both the amount of velocity dispersion as well as on the presence of substructure.
The interpretation of the core is in this case far from trivial.
According to \citet{Schulz+13}, it is due to the impossibility of accurately tracing the phase-space sheet \citep[similar to the discreteness effects advocated by][]{Binney04}, whereas \citet{Colombi&Touma14}, based on the waterbag method, point out that the constant-density core is actually a rather intricate and chaotic structure in phase space.
This structure is difficult to quantify because it does retain detailed memory of the initial conditions, but it is unrelated to discreteness effects, and it should be independent on the number of particles (in particular, it is expected to be present also in the limit $N\to\infty$).

All the three factors we just outlined (finite probability, discreteness, and small-scale fluctuations) are inherent to {the probabilistic} approach.
Each individual realisation of the initial condition samples the probability distribution $\Psi$ (which may be, in general, finite over all phase space) with a discrete number $N$ of particles.
Random initial conditions introduce a substantial amount of small-scale perturbations (Poisson noise), leading to the formation of numerous substructures and a complex (`chaotic') distribution of particles in phase space.
However, the details of such distribution are univocally determined by the particular realisation of the initial conditions, and they wash away after averaging over many independent realisations, eventually yielding a smooth probability distribution at all times.
This constitutes an important difference with respect to the traditional approach based on deterministic initial conditions, both for a finite $N$ as well as for a continuous fluid meant to represent $N\to\infty$.

Our results are consistent with~\cite{Binney04} in the sense that discreteness effects may lead to an overestimation of the core size in the cold-collapse (\emph{bar}) case {when $N_{\rm tracers}<N$}, but we would like to stress once again that this feature is indeed physical, and it will always be present in any real system with finite number of particles.
On the other hand, the core size tends to \emph{increase} with $N$ in the warm-collapse (\emph{box}) case.
Although {a more detailed study} would be required in order to test this conjecture, {our results show, consistent with previous findings \citep{JW10}, that systems with small $N$ evolve faster} towards thermal equilibrium, and therefore the solution is closer to the predictions of Maxwell-Boltzmann statistics (flatter than a power law, but more concentrated than constant $\Psi$).

{
In general, we find that the one-point probability distribution $\Psi$ in the quasi-steady state is consistent with the phenomenological analytical formulae that have previously been proposed to describe the distribution function $f$ and/or the coarse-grained phase-space density $\hat f$ in the deterministic interpretation.
Although this is, to some extent, expected, given that $\Psi$ (or equivalently, $\bar f$) is estimated from an average over individual N-body simulations, it is unclear that other results, such as e.g. the time required to attain the quasi-stationary state, are directly extrapolable (compare Figures~\ref{Fig_Psi_bar} and~\ref{Fig_individual}).
Moreover, we would like to argue that these formulae provide a valid, albeit partial and/or approximate description of the system.
They may be useful in practical terms, as well as a guideline towards a full physical understanding, but in our opinion only the latter can provide a complete and exact description, and therefore the details (e.g. the best-fitting value of the logarithmic slope of the density profile over a certain radial range) are less relevant than qualitative considerations, such as e.g. the asymptotic behaviour as $x$, $v$, and $\e$ go to zero or infinity (i.e. a constant phase-space density core at the centre and an exponential in $\e$ -- and thus $x$ and $v^2$ -- in the outskirts).
}

\subsection{{Insight on the three-dimensional case}}

{
The evolution of three-dimensional self-gravitating systems is much more complex than the academic case in one dimension that we have considered here \citep[see e.g.][and references therein]{Binney&Tremaine08, Levin+14}.
Most importantly, particle orbits may become unbound and escape the system, eventually leading to complete evaporation on a time scale proportional to the relaxation time~\eqref{eq_RelaxTime3D} rather than thermal equilibrium.
Dynamical friction, as well as two-body and collisional relaxation, are faster in three dimensions, and it is not obvious that our results can be easily extrapolated to real physical systems.
}

{
From a purely phenomenological point of view, the regularity of the light distribution in elliptical galaxies \citep{deVaucouleurs48} suggests that real physical systems do indeed reach a quasi-steady state with a `universal' structure.
This conjecture is also supported the by the results of cosmological N-body simulations, where it is well known that not only the radial density profile \citep{NFW97}, but most of the dynamical properties of dark matter haloes are `universal', at least in their central regions \citep[see e.g.][]{AG08}.
Moreover, the shape of the density profile (i.e. the internal structure of the quasi-stationary state) is also found do depend weakly on the number of particles, except for the size of the central core \citep{Power+03}.
}

{
If all our considerations were still valid in the three-dimensional case, one could argue, on the one hand, that the details of the initial conditions must play an important role on the precise shape of the quasi-steady state, and therefore} the small number of degrees of freedom observed in the dark matter haloes formed in cosmological  N-body simulations \citep[or even in the optical spectra of nearby galaxies, see e.g.][]{ASA11} merely reflects the statistical properties of Gaussian random peaks in the early universe \citep{A+04, A+07}.

{On the other hand, the relatively small dependence with $N$} hints that the phase-space structure of astrophysical N-body systems might be {different from} the predictions of the collisionless Boltzmann equation, strictly valid only for a continuous fluid.
A stellar cluster, for instance, would be an example of a warm system (starting from a turbulent molecular cloud) containing typically less than a million particles (stars), and therefore {we do expect} that finite-$N$ effects {play a} role in its evolution.
In contrast, even assuming a fairly heavy dark matter particle, with mass $m_{\rm dm}c^2\sim\mathcal{O}($TeV), the smallest dark matter haloes to host a galaxy would contain of the order of $N \sim \frac{ \rm 10^{8}\ M_\odot\ c^2 }{ \rm 1~Tev }\sim 10^{62}$ particles.
In this case, finding the exact time scale for diffusion in phase space is critical in order to gauge whether such system would be better described by the collisionless Boltzmann equation, by following the dynamics of ${N_{\rm tracers}}\sim\mathcal{O}(10^6)$~tracers, {or by developing a new methodology that explicitly accounts for $N$ (as well as other effects, such as e.g. the mixture of particle masses)}.

\section{Conclusions}
\label{sec_Conclusions}

In the present work we {discuss the merits and drawbacks of using a statistical ensemble of numerical $N$-body simulations to investigate} the one-point probability density $\Psi$ of finding a particle in a region $(x,v)$ of phase space{,} and the \emph{expected} phase-space density $\bar f(x,v,t)= M \Psi(x,v,t)$, where $M$ is the total mass of the system.
The difference with the {deterministic} approach {usually followed in astrophysics and cosmology} is both conceptual and practical.

On the one hand, {the statistical} interpretation is intrinsically discrete (the fine-grained distribution function $f$ for any individual system of $N$ particles is always a sum of Dirac deltas) and probabilistic (we are not interested on the evolution of one of these individual systems, but of a statistical ensemble of realisations of the probability density $\Psi$ at the initial time).
{This, of course, limits the range of problems for which the method will be feasible.
In particular, it would not be} suitable to study the evolution of a particular deterministic configuration (e.g. the stars in Orion), but of a generic class of systems with a similar distribution in phase space (e.g. stellar clusters of a given size).
{It would also be useful to investigate statistically-averaged properties such as e.g. the density profile or the velocity distribution function, but it is unclear to what extent (or how) it could be adapted to provide information on other quantities, such as, for instance, the amount and physical properties of substructures}.

On the other hand, the solution to the N-body problem under the {probabilistic} interpretation corresponds to an integral over all possible random realisations of the initial conditions.
In practice, an arbitrarily good approximation can be obtained by using a finite number $S$ of independent simulations with the same number of particles, where $\bar f$ and $\Psi$ can be trivially estimated from the number of particles inside a given bin in $(x,v)$ at time $t$.
At variance with traditional N-body simulations, the accuracy of the numerical solution can be improved by increasing the number of individual realisations.
From a computational point of view, {this} scheme has the {advantage} that increasing $S$ is much more efficient than increasing $N$ in terms of both memory and CPU time.

We have applied {such} method to two initial probability densities $\Psi(x,v,0)$: a \emph{bar} in phase space~\eqref{eq_barIC_Psi}, where particles are randomly distributed between $x=-\frac{1}{2}$ and $x=\frac{1}{2}$ with null initial velocity, and a \emph{box}~\eqref{eq_boxIC} where velocities are also random uniform between $v=-\frac{1}{2}$ and $v=\frac{1}{2}$.
Our results show that, analogously to the deterministic interpretation, $\Psi(x,v,t)$ evolves towards a quasi-stationary state that admits a description of the form $\Psi(\e)$, where $\e = \frac{1}{2} v^2 + \bar{\phi}(x)$ is the {expected} energy per unit of mass of a particle located at $(x,v)$, whose exact form keeps memory of the initial conditions {(i.e. it is \emph{not} `universal')} and depends slightly on the number of particles.
Most notably, the characteristic time to approach such state is an increasing function of $N$, and there are subtle changes on the size {and physical properties} of the central `core'.

We do argue that understanding the dependence with $N$ is of paramount importance to address the validity of using ${N_{\rm tracers}}\ll N$ tracers and to discriminate physical effects from numerical artefacts.
The {deterministic} interpretation often assumes that the physical evolution of real astrophysical systems is correctly described by the collisionless Boltzmann equation, and {studies} one particular initial configuration (e.g. a certain perturbation of equispaced particles) increasing $N$ as much as possible.
Our results suggest that finite-$N$ effects are in general not negligible in any real system; certainly not for $N\le 10^6$, and probably also for much larger numbers of particles.
Studying the evolution of $\Psi(x,v,t)$ as a function of $N$ is in our view a more adequate approach to understanding the N-body problem than focusing on the limit $N\to\infty$.

\section*{Acknowledgments}

{The authors are indebted to the referee, Stephane Colombi, for an extremely rigorous and constructive report, that has certainly helped to improve the clarity and accuracy of the manuscript}.
Financial support has been provided by projects AYA2013-47742-C4-3-P and AYA2016-79724-C4-1-P (\emph{Ministerio de Econom\'{i}a y Competitividad}; Mineco, Spain), as well as the exchange programme `Study of Emission-Line Galaxies with Integral-Field Spectroscopy' (SELGIFS, FP7-PEOPLE-2013-IRSES-612701), funded by the EU through the IRSES scheme.
YA is also supported by the \emph{Ram\'{o}n y Cajal} programme (RyC-2011-09461; Mineco, Spain).


\bibliographystyle{mnras}
\bibliography{references}

\label{lastpage}
\end{document}